\documentclass[11pt,twoside]{article}
\usepackage{asp2006}
\usepackage{epsfig}
\usepackage{epsf}
\usepackage{graphics}
\usepackage{lscape}
\bibpunct[, ]{(}{)}{;}{a}{}{,}

\def\edcomment#1{\iffalse\marginpar{\raggedright\sl#1\/}\else\relax\fi}
\reversemarginpar

\begin{document}
\title{Cosmic Evolution of Radio Sources in ATLAS}
\author{Minnie Y. Mao$^{1,2,3}$, Ray P. Norris$^{2}$, Rob Sharp$^{3}$, Jim E. J. Lovell$^{1}$}

\affil{$^{1}$School of Mathematics and Physics, University of Tasmania, Private Bag 37, Hobart, 7001, Australia\\
\vspace{0.2cm}
$^{2}$ CSIRO Australia Telescope National Facility, PO Box 76, Epping, NSW, 1710, Australia\\
\vspace{0.2cm}
$^{3}$ Anglo-Australian Observatory, PO Box 296, Epping, NSW, 1710, Australia}

\begin{abstract}
The Australia Telescope Large Area Survey (ATLAS), which is the widest deep field radio survey so far attempted, aims to probe the evolution of radio galaxies out to the edge of the Universe. Using AAOmega on the Anglo-Australian Telescope we have successfully obtained spectroscopic redshifts for 395 of the ATLAS radio galaxies. Coupled with 169 redshifts from the existing literature, we now have 564 spectroscopic redshifts for ATLAS sources. 
\end{abstract}

\vspace{-0.5cm}
\section{Introduction}

The Australia Telescope Large Area Survey (ATLAS) aims to image approximately seven square degrees of sky to 10 $\mu$Jy at 1.4 GHz, in two separate fields: Chandra Deep Field South (CDFS) and European Large Area ISO Survey-South 1 (ELAIS-S1). These fields were chosen to coincide with the \textit{Spitzer} Wide-Area Infrared Extragalactic (SWIRE) Survey program \citep{Lonsdale03} so that corresponding optical and infrared photometric data are available. CDFS also encompasses the Great Observatories Origins Deep Survey (GOODS) field \citep{Giavalisco04}. While ATLAS radio observations are only partially complete (the current rms noise is $\sim$20 - 30 $\mu$Jy), the first catalogues have been published \citep{Norris06,Middelberg08} yielding unexpected and interesting results \citep{Norris07}. There are currently $\sim$2000 radio sources detected in the ATLAS fields but we expect to obtain a sample of  $\sim$10000 at the completion of ATLAS.

The main goals of ATLAS are to understand the formation and evolution of galaxies as a function of cosmic time and determine the evolutionary relationship between AGNs and star-forming galaxies (SFGs) \citep{Norris06}. In order to do this we require distances to the radio sources, which in turn requires optical spectroscopy. Spectroscopy also yields line ratios which can differentiate between SFGs and AGN \citep{Baldwin81}, and between different classes of AGN \citep{Veilleux87, Kewley06, Randall09}. Photometric redshifts are unsuitable for this as  ATLAS sources frequently differ from the standard templates used for photometric fits. 

\section{Spectroscopic Data}

ATLAS sources were observed using the AAOmega spectrograph \citep{Sharp06} on the Anglo-Australian Telescope (AAT). The observations were undertaken in both standard multi-object spectroscopy (MOS) mode and nod+shuffle mode from December 1 to December 8, 2007. The sources were split into bright (R $<$ 21.5), intermediate (21.5 $<$ R $<$ 23) and nod+shuffle (R $>$ 21.5 but undetected in ``intermediate'') magnitude bins. We only obtained 22 hours of data because of poor weather thus we only partially completed the intermediate and nod+shuffle observations for CDFS and only observed the bright ELAIS sources.

We also observed a sample of non-radio sources from the CDFS field chosen on the basis that their 24 $\mu$m fluxes deviate from the radio-far-infrared correlation and are therefore unlikely to be star-forming galaxies. \citet{Norris09} have shown that these sources are probably obscured radio-quiet AGN. 

\section{Results}

Our AAT observations yielded a total of 395 new spectroscopic redshifts for the ATLAS sources and 489 new spectroscopic redshifts for the 24 $\mu$m sources. Coupled with redshifts from the literature there are now 564 ATLAS sources with spectroscopic redshifts, summarized in Table \ref{summary}.  Table \ref{obsresults} shows the detection rates for both the ATLAS radio sources and the 24 $\mu$m sources.

\begin{table}
\begin{center}
\small{
\caption{Spectroscopic redshifts available for ATLAS sources. Column 1 gives the name of the ATLAS field. Column 2 gives the total number of ATLAS sources while Columns 3 and 4 give the number of spectroscopic redshifts obtained from the AAT and the literature respectively. Column 5 gives the total number of available spectroscopic redshifts.}\label{summary}
\begin{tabular}{lrrrr}
\\
\textbf{Field} & \textbf{Total No.} & \textbf{AAT} & \textbf{Literature} & \textbf{Total with zsp} \\
CDFS & 728 & 143 & 118 & 261\\
ELAIS & 1276 & 252 & 51 & 303\\
Total & 2004 & 395 & 169 & 564\\

\end{tabular}
}
\end{center}
\end{table}

\vspace{-1cm}

\begin{table}
\begin{center}
\small{
\caption{Success rates for the Dec 2007 AAOmega observations. Column 1 gives the name of the field while Columns 2, 3 and 4 show the successful redshift detection rate for all targeted sources, the ATLAS radio sources and the 24 $\mu$m sources respectively.}\label{obsresults}
\begin{tabular}{lrrr}
\\

\textbf{Field} & \textbf{ATLAS + 24 $\mu$m} & \textbf{ATLAS}& \textbf{24 $\mu$m}\\
CDFS bright & 63\% & 42\% & 94\%\\
CDFS intermediate & 60\% & 22\% & 76\%\\
CDFS N+S & 30\% & 15\% & 37\% \\
\end{tabular}
}
\end{center}
\end{table}

\begin{figure}[h]
\begin{center}
\includegraphics*[angle=-90, scale=0.3]{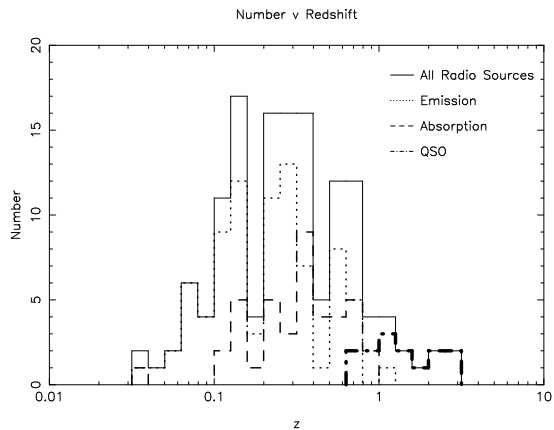} 
\end{center}
\small{
\caption{Histogram of spectroscopic redshifts plotted logarithmically for CDFS. The different line styles represent the different types of spectra (emission, absorption or QSO). Redshift data for the 143 sources shown in this plot were taken from AAOmega observations in December 2007 \citep{Mao09}.}\label{zhist}
}
\end{figure}

Figure \ref{zhist} plots the redshift distribution for the ATLAS sources in CDFS. The relatively larger number of QSOs at z $>$ 1 is most likely due to selection effects such as Malmquist Bias. Furthermore, the greater number of detected emission line spectra may be due to emission lines being easier to identify in low signal-to-noise data than absorption lines.

\section{Discussion and Conclusions}
Currently we have spectroscopic redshifts for over a quarter of all ATLAS sources. In order to obtain a more complete redshift catalogue it is necessary to obtain time on larger telescopes such as Gemini so that we may probe the fainter sources and hence understand the incompleteness at fainter magnitudes. 

The detection rate for the 24 $\mu$m sources was significantly higher than for the ATLAS radio sources, suggesting they have stronger emission lines, although their selection implies that they are more likely to be AGN, which generally have weaker emission lines and a lower redshift detection rate. Understanding this paradox is likely to have significant implications for our understanding of early galaxy evolution. It is only through studies like ATLAS that we may build up a comprehensive picture of the cosmic evolution of radio galaxies.

\end{document}